\documentclass{aa} 
\usepackage{graphicx} 
\begin{document}

\title{Oxygen and nitrogen abundances in Virgo and field spirals}

\author{L.S.~Pilyugin \inst{1},  
        Mercedes Moll\'{a} \inst{2},
        Federico Ferrini \inst{3,}\inst{4},
        Jos\'{e} M. V\'{\i}lchez \inst{5} }

\offprints{L.S.~Pilyugin }

   \institute{   Main Astronomical Observatory
                 of National Academy of Sciences of Ukraine,
                 27 Zabolotnogo str., 03680 Kiev, Ukraine, \\
                 (pilyugin@mao.kiev.ua)
                 \and
                 Departamento de F\'{\i}sica Te\'{o}rica, 
                 Universidad Aut\'onoma de Madrid, 
                 28049 Cantoblanco, Spain, \\
                 (mercedes@pollux.ft.uam.es)
                  \and
                 Department of Physics, Section of Astronomy,
                 University of Pisa, piazza Torricelli 2,
                 56100 Pisa, Italy,    
                  \and
                 INTAS, 58 Avenue des Arts, 1000 Bruxelles, Belgium \\
                 (ferrini@intas.be) 
                 \and
                 Instituto de Astrof\'{\i}sica de Andaluc\'{\i}a,
                 Apdo. 3004, 18080 Granada, Spain  \\
                 (jvm@iaa.es)
                 }

\date{Received 20 September 2001 / Accepted 6 November 2001}

\abstract{The oxygen and nitrogen abundances in the H\,{\sc ii} regions of the 
nine Virgo spirals of the sample from \cite{ski96} and in nine field spiral 
galaxies are re-determined with the recently suggested P -- method. We confirm 
that there is an abundance segregation in the sample of Virgo spirals in the 
sense that the H\,{\sc i} deficient Virgo spirals near the core of the cluster 
have higher oxygen abundances in comparison to the spirals at the periphery of 
the Virgo cluster. At the same time both the Virgo periphery and core spirals 
have counterparts among field spirals. Some field spirals have H\,{\sc i} to 
optical radius ratios, similar to that in H\,{\sc i} deficient Virgo core 
spirals. We conclude that if there is a difference in the abundance properties 
of the Virgo and field spirals, this difference appears to be small and masked 
by the observational errors.  
}

\titlerunning{Oxygen and nitrogen abundances in Virgo and field spirals}

\authorrunning{L.S.~Pilyugin et al.}  

\maketitle

\section{Introduction}

There is evidence that the environment affects the properties of
galaxies in clusters (see review of \cite{bal92}). The most obvious
effect is probably that concerning the H\,{\sc i} content of
galaxies in clusters (\cite{hay86,huc89,cay94}).  It is well 
established that the spiral galaxies of the Virgo cluster have
a tendency to be H\,{\sc i} deficient in comparison with normal field
spirals (Solanes et al. 1996, 2001) and this deficiency is correlated with
distance to the cluster center, the proportion of gas-poor spirals
increasing continuously towards the cluster center. Thus, a cluster
galaxy (especially a galaxy near the center of the cluster) evolves in a
different surrounding gaseous medium in comparison to a field galaxy.

The gas exchange between a galaxy and its ambient medium (loss of
gas by the galaxy or gas infall onto the galaxy) changes the course of
the chemical evolution of the galaxy. Then, it can be expected that
the environment affects the chemical properties of galaxies in
clusters.  Oxygen plays a key role in understanding the (chemical)
evolution of galaxies.  The origin of oxygen seems to be reliably
established in contrast to other elements like carbon or nitrogen. The
oxygen abundance can be considered as a tool to investigate the
evolution of galaxies. For example, the value of oxygen abundance in a
galaxy combined with the value of the gas mass fraction can tell us
about the efficiency of mass exchange between a galaxy and its
environment (Pilyugin \& Ferrini 1998, 2000).

A number of works have been devoted to searching for the possible
effects of cluster environment on the chemical properties of spiral
galaxies (\cite{shi91}; Henry et al. 1992, 1994, 1996; \cite{ski96}).  
The conclusions of
these works are: {\it i)} the spirals at the periphery of the cluster
are indistinguishable from the field galaxies, {\it ii)} the H\,{\sc
i} deficient Virgo core galaxies have larger oxygen abundances at a
predetermined galactocentric distance $r = 0.4R_{25}$ (where R$_{25}$
is the isophotal radius) than the field galaxies of comparable luminosity
or Hubble type.

Accurate oxygen abundances can be derived from measurement of
temperature-sensitive line ratios, such as
[OIII]4959,5007/[OIII]4363. This method will be referred to as the
T$_{e}$ - method. Unfortunately, in oxygen-rich H\,{\sc ii} regions
the temperature-sensitive lines such as [OIII]4363 are too weak to be
detected. For such H\,{\sc ii} regions, empirical abundance indicators
based on more readily observable lines were some years ago suggested
(\cite{pag79,allo79}).  The empirical oxygen abundance indicator
R$_{23}$ = ([OII]3727,3729 + [OIII]4959,5007)/H$_{\beta}$, suggested
by \cite{pag79}, has found widespread acceptance and it has been used
for the oxygen abundance determination in H\,{\sc ii} regions where
the temperature-sensitive lines are undetectable. This method will be
referred to as the R$_{23}$ - method. Using the R$_{23}$ - method, the
characteristic oxygen abundances (the oxygen abundance at a
predetermined galactocentric distance) and radial oxygen abundance
gradients were obtained for a large sample of field spiral galaxies
(\cite{vil92,zar94,vzee98}, among others).

However, the basic problem whether R$_{23}$ is an accurate abundance
indicator is open to discussion (\cite{zar92,kin94}, among others). It
has been found (\cite{pil00}) that the error in the oxygen abundance
derived with the R$_{23}$ -- method involves two parts: the first is a
random error and the second is a systematic error depending on the
excitation parameter. A new way of oxygen abundance determination in
H\,{\sc ii} regions (P -- method) has been recently suggested
(Pilyugin 2000, 2001a). By comparing oxygen abundances derived through the
$T_{e}$ --  method in high-metallicity H\,{\sc ii} regions with those
derived through the P -- method, it has been found that the precision of
oxygen abundance determination with the P -- method is comparable to
that of the $T_{e}$ -- method (Pilyugin 2001a,b). It has been also
shown that the R$_{23}$ -- method provides more or less realistic
oxygen abundances in high-excitation H\,{\sc ii} regions but it
produces overestimated oxygen abundances in low-excitation H\,{\sc ii}
regions. Taking into account this fact together with the fact known
for a long time (\cite{sea71,smi75}) that galaxies can show strong
radial excitation gradients, in the sense that only the low-excitation
H\,{\sc ii} regions populate the central parts of some galaxies, one
can expect that the oxygen abundances of the inner H\,{\sc ii} regions
and the gradient slopes based on the (O/H)$_{R_{23}}$ data 
can be appreciably overestimated.
This speculation has been confirmed by comparison of the radial
(O/H)$_{R_{23}}$ abundance distribution with the radial
(O/H)$_{T_{e}}$ abundance distribution within the disk of the
well-observed spiral galaxy M101 (\cite{pil01b}).

The problem whether the cluster environment affects the chemical
evolution of galaxies will be considered here based on the (O/H)$_{P}$
and (N/H)$_{P}$ abundances obtained in Section 2. In \S 3 we analyze
these data with the multiphase model applied to three Virgo
galaxies, considered as typical examples. A dicussion is included
in Section 4 and Conclusions are in Section 5.

\section{Virgo and field spiral abundances}

\subsection{Oxygen}

%++++++++++++++++++++++++++++++++++++++++++++++++++++++ Table sample
\begin{table*}
\caption[]{\label{table:sample}
The galaxy sample.
}
\begin{center}
\begin{tabular}{lclrcccclcl} \hline \hline
           &           &            &          &                &          &\multicolumn{2}{c}{P method}       &                          &\multicolumn{2}{c}{R$_{23}$ method}\\  \cline{7-8} \cline{10-11}
Galaxy     &  d        &  ref       &logL$_B$  & R$ _{25}$      & T type   &  O/H$_0$  & gradient              & literature sources       &     O/H$_0$  &   gradient      \\  
           &  Mpc      &            &          &     (')        &          &           & dex/R$ _{25}$         & for the  spectra         &              &   dex/R$ _{25}$ \\  
           &           &            &          &                &          &           &                       &                          &              &                 \\   \hline
           &           &            &          &                &          &           &                       &                          &              &                 \\   
\multicolumn{11}{c}{Virgo cluster galaxies}                                                                                                                                \\
           &           &            &          &                &          &           &                       &                          &              &                 \\   
NGC~4254   &  16.8     & SKSZ       &  10.61   &     2.81       &    5     &   8.94    &   -0.65               &  HPC,MRS,SSK             &    9.44      &   -0.65         \\   
NGC~4303   &  16.8     & SKSZ       &  10.62   &     3.23       &    4     &   8.84    &   -0.72               & HPLC,SSK                 &    9.43      &   -1.05         \\   
NGC~4321   &  16.8     & SKSZ       &  10.66   &     3.79       &    4     &   8.86    &   -0.37               & MRS,SSK                  &    9.36      &   -0.38         \\   
NGC~4501   &  16.8     & SKSZ       &  10.70   &     3.54       &    3     &   8.99    &   -0.52               & SKSZ                     &    9.48      &   -0.45         \\   
NGC~4571   &  16.8     & SKSZ       &   9.94   &     1.86       &    6     &   8.90    &   -0.20               & SKSZ,SSK                 &    9.34      &   -0.10         \\   
NGC~4651   &  16.8     & SKSZ       &  10.24   &     2.04       &    5     &   8.71    &   -0.64               & SKSZ                     &    9.25      &   -0.85         \\   
NGC~4654   &  16.8     & SKSZ       &  10.34   &     2.45       &    6     &   8.84    &   -0.77               & SKSZ                     &    9.34      &   -0.84         \\   
NGC~4689   &  16.8     & SKSZ       &  10.10   &     2.18       &    4     &   8.88    &   -0.41               & SKSZ                     &    9.40      &   -0.39         \\   
NGC~4713   &  16.8     & SKSZ       &   9.88   &     1.35       &    7     &   8.71    &   -0.73               & SKSZ                     &    9.24      &   -1.02         \\   
           &           &            &          &                &          &           &                       &                          &              &                 \\   
\multicolumn{11}{c}{Field galaxies}                                                                                                                                        \\
           &           &            &          &                &          &           &                       &                          &              &                 \\   
NGC~628    &   9.7     & ZKH        &  10.25   &     5.36       &    5     &   8.71    &  --0.48               & MRS, vZ, FGW             &    9.31      &  --0.77         \\   
NGC~1232   &  21.5     & vZ         &  10.69   &     3.71       &    5     &   8.73    &  --0.58               &     vZ                   &    9.36      &  --0.88         \\   
NGC~2903   &   8.9     & DK         &  10.45   &     6.29       &    4     &   8.94    &  --0.73               & MRS, ZKH, vZ             &    9.44      &  --0.76         \\   
NGC~3184   &   8.7     & vZ         &   9.93   &     3.71       &    6     &   8.97    &  --0.63               & MRS, ZKH, vZ             &    9.53      &  --0.80         \\   
NGC~3351   &   8.1     & ZKH        &   9.90   &     3.79       &    3     &   8.94    &  --0.31               & MRS, OK, FGW             &    9.43      &  --0.33         \\   
NGC~5194   &   7.7     & ZKH        &  10.49   &     5.61       &    4     &   8.92    &  --0.40               & MRS, D, BKG              &    9.43      &  --0.45         \\   
NGC~5236   &   4.5     & S          &  10.31   &     6.59       &    5     &   8.75    &  --0.24               & DTJS, WS                 &    9.26      &  --0.27         \\   
NGC~5457   &   7.5     & KG,vZ      &  10.67   &    14.42       &    6     &   8.76    &  --0.77               &  KG                      &    9.37      &  --1.52         \\ 
NGC~6946   &   5.9     & KSH        &  10.60   &     8.30       &    6     &   8.73    &  --0.49               & MRS, FGW                 &    9.24      &  --0.62         \\   \hline  \hline 
\end{tabular}
\end{center}

\vspace{0.05cm}

{\it List of references}:

BKG   -- \cite{bre99};
D     -- \cite{di91};
DK    -- \cite{dro00};
DTJS  -- \cite{duf80};
FGW   -- \cite{fer98};
HPC   -- \cite{hen94};
HPLC  -- \cite{hen92};
KG    -- \cite{ken96};
KSH   -- \cite{kar00};
MRS   -- \cite{mca85};
OK    -- \cite{oey93};
S     -- \cite{sch94};
SKSZ  -- \cite{ski96};
SSK   -- \cite{shi91};
vZ    -- \cite{vzee98};
WS    -- \cite{web83};
ZKH   -- \cite{zar94};
\end{table*}

The (O/H)$_{P}$ oxygen abundances in H\,{\sc ii} regions are
determined with the expression suggested in (\cite{pil01a})
\begin{equation}
12+log(O/H)_{P} = \frac{R_{23} + 54.2  + 59.45 P + 7.31 P^{2}}
                       {6.07  + 6.71 P + 0.37 P^{2} + 0.243 R_{23}}  ,
\label{equation:ohp}
\end{equation}
where $R_{23}$ =$R_{2}$ + $R_{3}$, 
$R_{2}$ = $I_{[OII] \lambda 3727+ \lambda 3729} /I_{H\beta }$, 
$R_{3}$ = $I_{[OIII] \lambda 4959+ \lambda 5007} /I_{H\beta }$, 
and P = $R_{3}$/$R_{23}$. 

For comparison, the (O/H)$_{R_{23}}$ oxygen abundances are also
determined.  Several workers have suggested calibrations of the
R$_{23}$ in terms of the oxygen abundance 
(\cite{edm84,mca85,dop86,zar94}, among others).  Abundances 
for a large sample of
spiral galaxies were derived by Zaritsky et al. (1994). The Zaritsky et al's
calibration is an average of the three calibrations by
\cite{edm84,mca85}, and \cite{dop86}. The oxygen abundances in Virgo
spiral galaxies were determined by Skillman et al. (1996) using this same
empirical R$_{23}$ -- abundance calibration.  Then this latter R$_{23}$
calibration has been adopted here for the determination of the
(O/H)$_{R_{23}}$ oxygen abundances.

The set of available spectra from the literature has been used for the
determination of the (O/H)$_{P}$ and (O/H)$_{R_{23}}$ oxygen
abundances. There exist nine Virgo spirals with line intensity
measurements for at least four H\,{\sc ii} regions (\cite{ski96}). We
will compare the abundance properties of this Virgo sample to a sample
of field spiral galaxies.  This sample of field spiral galaxies also
consists of nine galaxies, which were chosen in such way that {\it 1)}
they lie in the same range of luminosity as the Virgo sample spirals,
{\it 2)} they cover the same range of morphological type as the Virgo
spirals, and {\it 3)} they have large enough number of H\,{\sc ii} regions 
with measurements of oxygen and nitrogen lines. The adopted and computed
parameters of Virgo and field spiral galaxies are summarized in Table
\ref{table:sample}. The NGC number is listed in column 1. The adopted
distance is reported in column 2 (source for the distance in column
3).  The blue luminosity of the galaxy is reported in column 4.  The
isophotal radius R$_{25}$ in arcmin and the numerical Hubble type of
the galaxy (T type) taken from \cite{rc3} (RC3) are given in columns
5 and 6, correspondingly.  The central (O/H)$_{P}$ oxygen abundance
and the gradient expressed in terms of dex/R$_{25}$ are listed in
columns 7 and 8.  The source(s) for the line intensity measurements is
reported in column 9.  The central (O/H)$_{R_{23}}$ oxygen abundance
and the gradient expressed in terms of dex/R$_{25}$ are listed in
columns 10 and 11, respectively.
 
%============================================================Fig 01 deva
\begin{figure*}
\centering \includegraphics[width=17.0cm,angle=0]{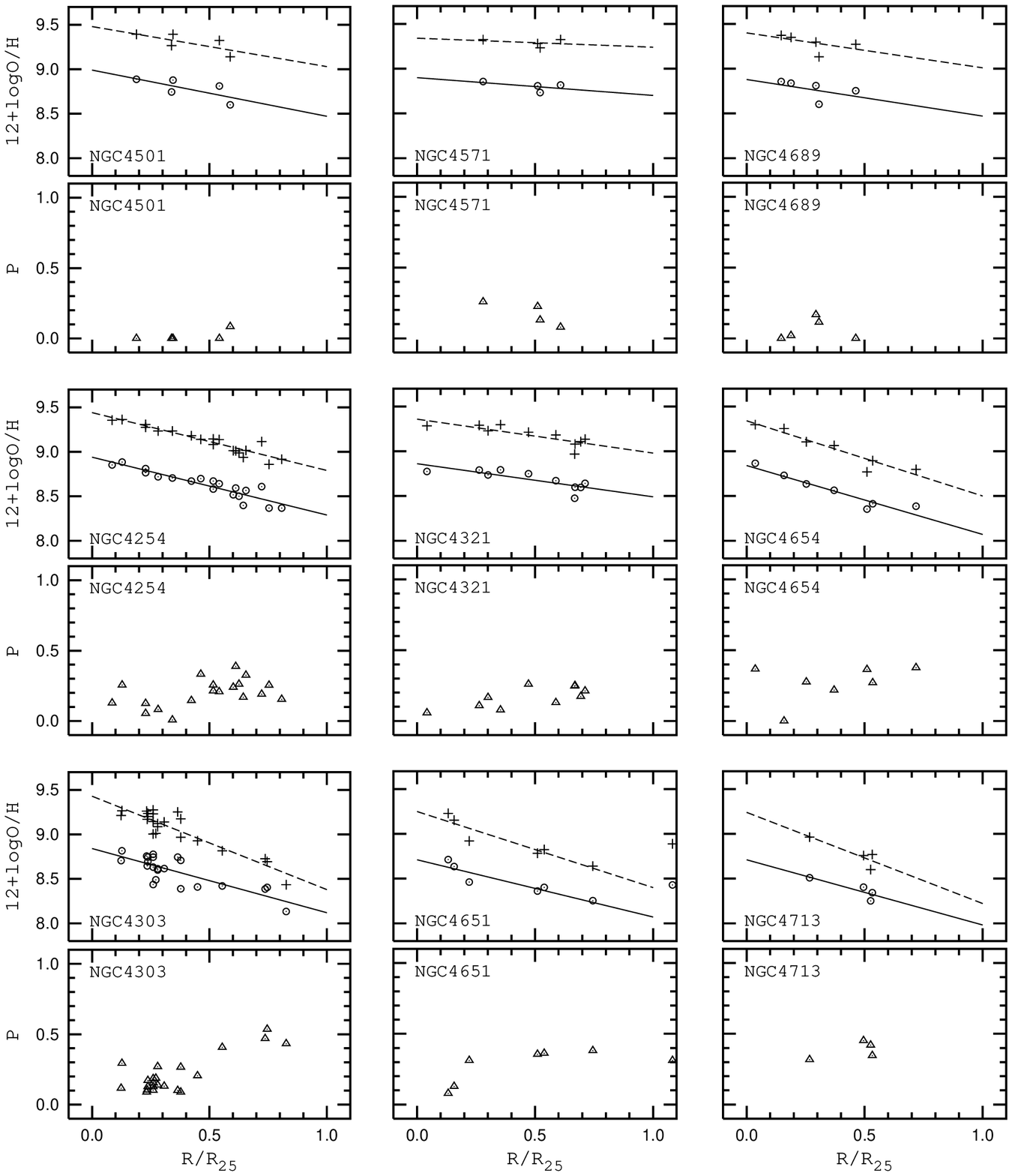}
\caption{ Oxygen abundances and values of excitation parameter P
versus galactocentric distance for Virgo spiral galaxies. The oxygen
abundances determined with the R$_{23}$ - method calibrated by
\cite{zar94} are shown by pluses, the best fits to these data
are presented by dashed lines. The oxygen abundances determined with
the P - method calibrated by \cite{pil01a} are shown by circles,
the best fits to these data are presented by solid lines. The values
of excitation parameter P are shown by triangles. The galactocentric
distances are normalized to the isophotal radius.  }
\label{figure:deva}
\end{figure*}

Fig.\ref{figure:deva} shows the derived (O/H)$_{P}$ and
(O/H)$_{R_{23}}$ oxygen abundances and excitation parameter P for
H\,{\sc ii} regions in Virgo spirals as a function of galactocentric
distance (normalized to the isophotal radius).  The radial
(O/H)$_{R_{23}}$ gradients in the Virgo spiral galaxies derived here
are close to those determined by Skillman et al. (1996), with the exception of
the galaxy NGC~4651. For this galaxy, those authors have found a value 
of the gradient of --0.42 (dex/R$_{25}$), while a value --0.85 (dex/R$_{25}$) 
is obtained here. The H\,{\sc ii} region +131,+021 with
galactocentric distance 1.08R$_{25}$ is responsible for the small
value of the gradient derived by those authors for that galaxy. Close 
examination of the radial distribution of (O/H)$_{P}$ oxygen abundance
within the disk of NGC~4651 (Fig.\ref{figure:deva}) shows that the
oxygen abundance at galactocentric distances larger than $\sim$
0.8R$_{25}$ is expected to be below  12+log(O/H) = 8.2.  Therefore it is 
unlikely that the H\,{\sc ii} regions within the disk of NGC~4651 at 
galactocentric distances larger than $\sim$ 0.8R$_{25}$  belong to the upper
branch of the O/H -- R$_{23}$ diagram and, as consequence, the
R$_{23}$ calibrations of the upper branch of the O/H -- R$_{23}$ diagram
should not be used for the oxygen abundance determination in the H\,{\sc
ii} region +131,+021.  Therefore this region has been excluded from
consideration here, resulting in a larger value of the gradient as
compared to the previous value of the gradient in Skillman et al. (1996).

As we can see in Fig.\ref{figure:deva} most of the H\,{\sc ii} regions
observed in the Virgo spiral galaxies are of low-excitation and they
do not show (or show marginal) radial excitation
gradients. Therefore  the (O/H)$_{P}$ gradients are close
to the (O/H)$_{R_{23}}$ gradients, but the abundance
(O/H)$_{P}$ at any given galactocentric distance is significantly
lower than the value estimated from the (O/H)$_{R_{23}}$ calibration.
This reflects the fact that the R$_{23}$ -- method may provide more or less 
realistic oxygen abundances for high-excitation H\,{\sc ii}
regions but produce overestimated oxygen abundances in low-excitation
H\,{\sc ii} regions. As an example, we see that a more or less distinct
radial excitation gradient can be seen in the Virgo spiral
NGC~4303. In consequence the (O/H)$_{P}$ gradient in the NGC~4303 is
steeper than the (O/H)$_{R_{23}}$ gradient, Fig.\ref{figure:deva}.

%============================================================Fig 02 pole
\begin{figure*}
\centering \includegraphics[width=17cm,angle=0]{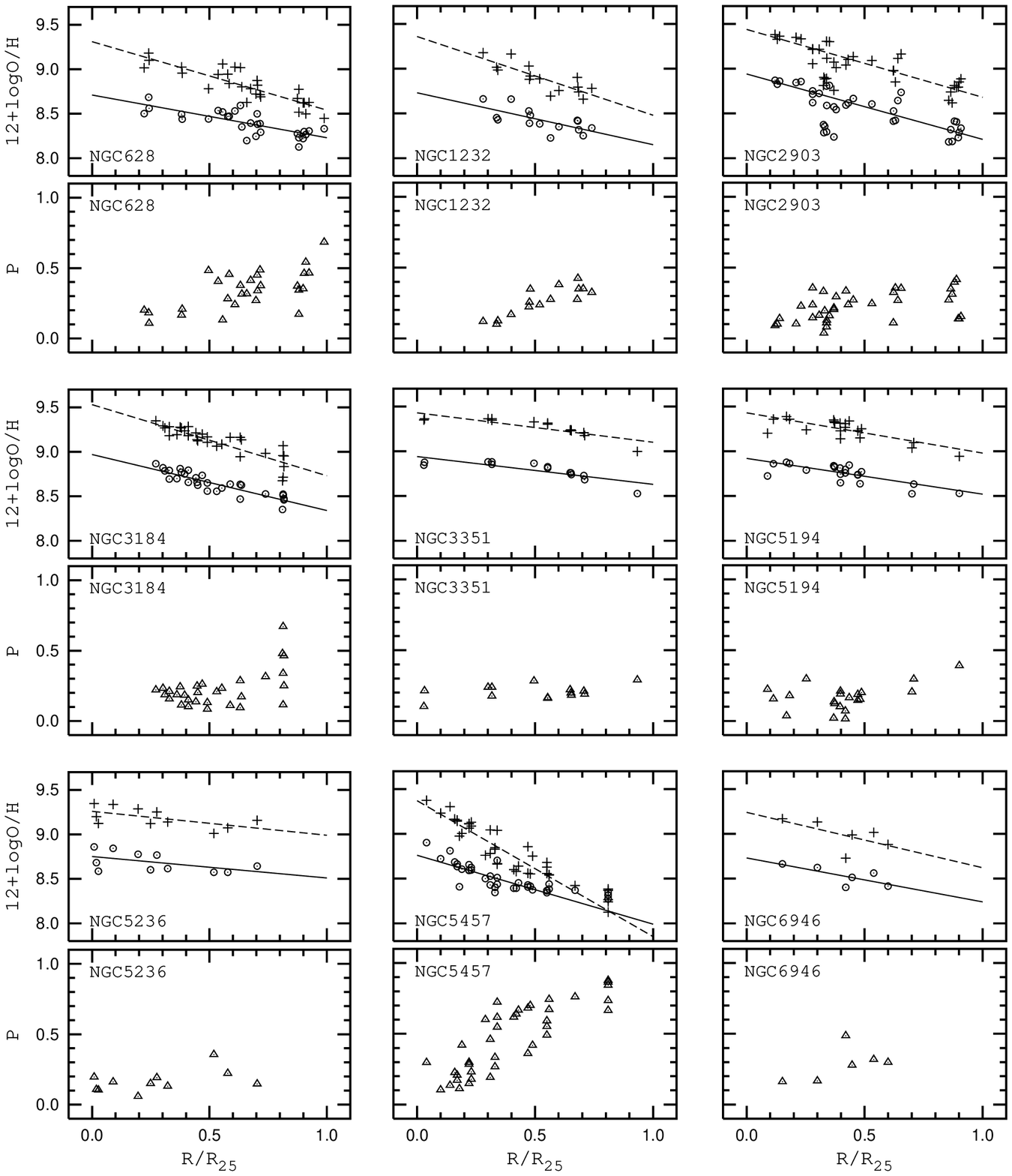}
\caption{ The same than Fig.\ref{figure:deva} for the field spiral
galaxies. }
\label{figure:pole}
\end{figure*}

Fig.\ref{figure:pole} shows the derived (O/H)$_{P}$ and
(O/H)$_{R_{23}}$ oxygen abundances and excitation parameter P for
H\,{\sc ii} regions in field spiral galaxies.  Some of them (NGC~628,
NGC~1232, NGC5457) show an appreciable radial excitation gradients and,
as a consequence, the (O/H)$_{P}$ gradients in these galaxies are
steeper than the (O/H)$_{R_{23}}$ gradients, Fig.\ref{figure:pole}.

%============================================================Fig 03 l
\begin{figure}
\resizebox{\hsize}{!}{\includegraphics[angle=0]{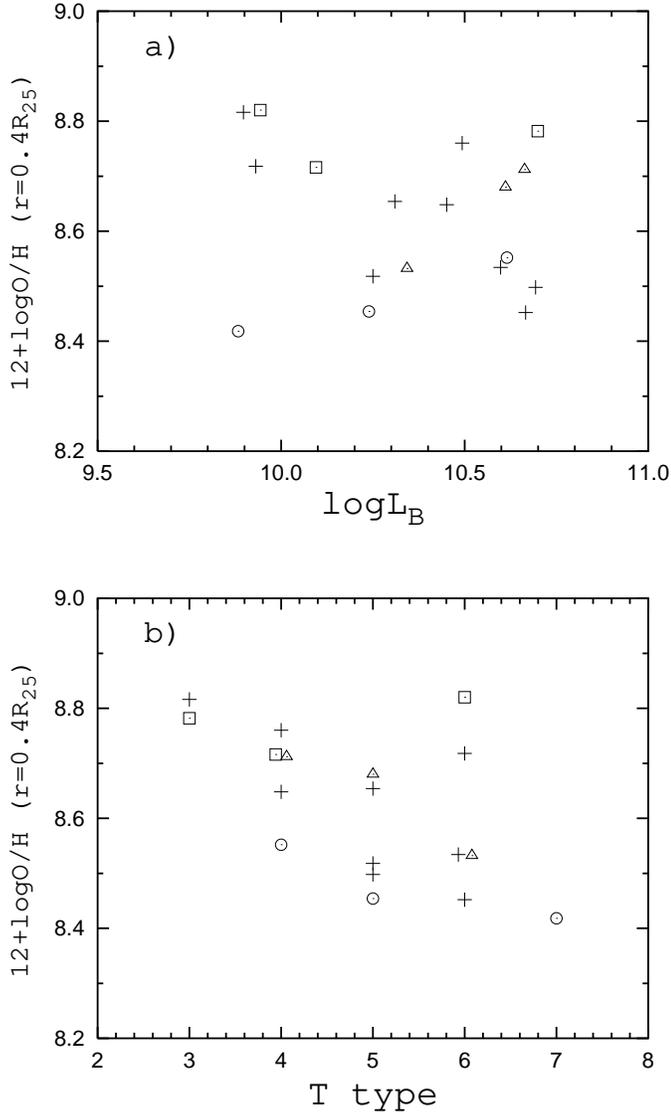}}
\caption{ Oxygen abundance at r=0.4~R$_{25}$ as a function of blue
luminosity {\bf (a)} and as a function of T type {\bf (b)} for the
Virgo and field spirals.  The pluses correspond to the field galaxies
(NGC~628, NGC~1232, NGC~2903, NGC~3184, NGC~3351, NGC~5194, NGC~5236,
NGC~5457, NGC~6946), the
squares represent the H\,{\sc i} deficient (Virgo core) spirals (NGC~4501,
NGC~4571, NGC~4689), the triangles represent the
intermediate galaxies (NGC~4254, NGC~4321, NGC~4654), and the open
circles represent the Virgo galaxies with normal H\,{\sc i} disks (NGC~4303,
NGC~4651, NGC~4713).  }
\label{figure:l}
\end{figure}

Fig.\ref{figure:l} shows oxygen abundances at r=0.4~R$_{25}$ as a
function of blue luminosity {\bf (a)} and as a function of the
morphological type T {\bf (b)} for the Virgo and field spirals. The
oxygen abundances at r=0.4~R$_{25}$ have been used instead of the central 
oxygen abundances because the central oxygen abundances in these
galaxies can be affected by large uncertainties: the number of H\,{\sc ii}
regions in the Virgo core galaxies is low and they cover a
small interval of galactocentric distances, see Fig.\ref{figure:deva}, 
implying that the extrapolation method used to estimate these central
data is uncertain. Since the value of r=0.4R$_{25}$ is within the
interval of galactocentric distances covered by the observed H\,{\sc
ii} regions in each galaxy, we believe the oxygen abundance at 
r=0.4R$_{25}$ is more reliable than the central oxygen abundance.

Skillman et al. (1996) divided the nine Virgo spirals in 
their sample into three groups of three galaxies each: 
those with strong H\,{\sc i} deficiencies (NGC4501, NGC4571, and NGC4689), 
intermediate cases (NGC4254, NGC4321, and NGC4654), and those 
with no H\,{\sc i} deficiencies (NGC4303, NGC4651, and NGC4713).  
The three subgroups into which the sample has been divided are 
coded with different symbols in Fig.\ref{figure:l}.  
The pluses correspond to the field galaxies, the squares represent
the H\,{\sc i} deficient (Virgo core) spirals, the triangles 
correspond to the intermediate galaxies, and the open circles
represent the Virgo galaxies with normal H\,{\sc i} disks.

%============================================================Fig 04 dh
\begin{figure}
\resizebox{\hsize}{!}{\includegraphics[angle=0]{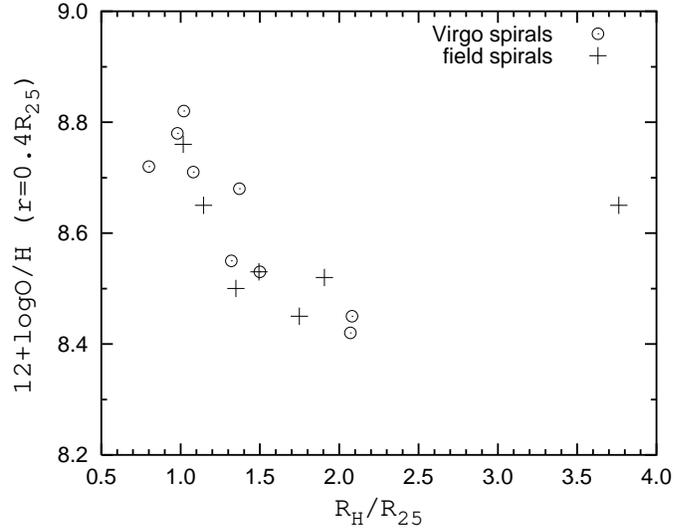}}
\caption{
12+logO/H at r=0.4R$_{25}$ versus R$_H$/R$_{25}$ diagram for Virgo (circles) 
and field (pluses) spiral galaxies. }
\label{figure:dh}
\end{figure}

The examination of Figs.\ref{figure:l}a,b shows that there is an
abundance segregation in the sample of Virgo spirals in the sense
that the H\,{\sc i} deficient Virgo core galaxies are all near the
top of the abundance distributions of galaxies with similar
properties, such as it was obtained by Skillman et al. (1996).  At the same
time, some field spiral galaxies have oxygen abundances similar to
those ones from the most oxygen-rich (H\,{\sc i} deficient) Virgo
spirals.

Those authors also found that the characteristic oxygen abundance (oxygen
abundance at r=0.4~R$_{25}$) for the Virgo cluster galaxies correlates
with the H\,{\sc i} to optical radius ratio, R$_H$/R$_{25}$, where
R$_H$ is the H\,{\sc i} radius at the isophotal level of 1 M$_{\sun}$
pc$^{-2}$ after Warmels (1988). Our O/H abundances at r=0.4R$_{25}$
{\sl versus} R$_H$/R$_{25}$ for Virgo spiral galaxies are shown in
Fig.\ref{figure:dh} with circles. A trend of decreasing characteristic
O/H value with increasing R$_H$/R$_{25}$ is evident for these
galaxies. We also derive the H\,{\sc i} to optical radius ratio for
the field spiral galaxies for which 21~cm observations are
available. We find R$_H$/R$_{25} =$ 1.91 for NGC~628, 1.35 for
NGC~1232, 1.14 for NGC~2903, 1.02 for NGC~5194, 3.76 for NGC~5236 ,
1.75 for NGC~5457 and 1.49 for NGC~6946, with 21~cm data from
Shostak \& van der Kruit (1984), van Zee \& Bryant (1999), Wevers 
et al. (1986), Shane (1975), Huchtmeier \& Bohnenstengel (1981), 
Boulanger \& Viallefond (1992).  Our O/H abundances at
r=0.4R$_{25}$ versus R$_H$/R$_{25}$ for field spiral galaxies are
shown in Fig.\ref{figure:dh} with pluses. Fig.\ref{figure:dh} shows that 
some field spirals have a small H\,{\sc i} to optical radius ratios, 
similar to that in H\,{\sc i} deficient Virgo core spirals. 
The field galaxies from our limited sample, with the exception of NGC~5236, 
show a similar trend of decreasing characteristic O/H value with increasing
R$_H$/R$_{25}$ as the Virgo spiral galaxies. 

Thus, the consideration of the O/H$_{P}$ oxygen abundances in H\,{\sc ii} 
regions in the Virgo and field spiral galaxies suggests that 
in general the Virgo spirals are indistinguishable from the field
spirals, both the Virgo periphery and core spirals have counterparts
among field spirals. 
There are no H\,{\sc i} deficient Virgo spirals with low oxygen
abundances.  This fact together with the larger number of H\,{\sc i}
deficient galaxies in the core of the cluster causes the apparent
segregation of high abundances in the center galaxies.

\subsection{Nitrogen}

The N/O abundance ratio in H\,{\sc ii} regions have been determined using
 algorithms following below. It was adopted  
\begin{equation}
\frac{N}{O} = \frac{N^+}{O^+}
\label{equation:nono}
\end{equation}
The $\frac{N^+}{O^+}$ value is derived from the expression (Pagel et al. 1992)
\begin{eqnarray}
\log (N^+/O^+) = \log \frac{I_{[NII] \lambda 6548 + \lambda 6584}}
{I_{[OII] \lambda 3726 + \lambda 3729}} + 0.307 - \frac{0.726}{t_2}  
\nonumber  \\ 
 + 0.40 \log t_2 - \log (1+1.35 x)
\label{equation:no}
\end{eqnarray}
where 
\begin{equation}
x= 10^{-4} n_e t_2^{-1/2}, 
\end{equation}
and $n_e$ is the electron density in cm$^{-3}$, $t_2$ = $t_{[NII]}$ is the 
electron temperature in units of 10$^4$K. 
It was adopted $t_2$ = $t_{[NII]}$  = $t_{[OII]}$. 
The $t_2$ value can be determined from the following equations:
\begin{eqnarray}
12+ \log (O^{++}/H^+) = \log \frac{I_{[OIII] \lambda 4959 + \lambda 5007}}
{I_{H_{\beta}}} + 6.174 +  \nonumber  \\ 
\frac{1.251}{t_3}  - 0.55 \log t_3 ,
\label{equation:oplus2}
\end{eqnarray}
\begin{eqnarray}
12+ \log (O^{+}/H^+) = \log \frac{I_{[OII] \lambda 3726 + \lambda 3729}}
{I_{H_{\beta}}} + 5.890 +  \nonumber  \\ 
\frac{1.676}{t_2}  - 0.40 \log t_2 + \log (1+1.35x)  ,
\label{equation:oplus}
\end{eqnarray}
\begin{equation}
\frac{O}{H} = \frac{O^+}{H^+} + \frac{O^{++}}{H^+}                ,
\label{equation:otot}
\end{equation}
\begin{equation}
t_2 =  0.7 \, t_3 + 0.3 .
\label{equation:t2t3}
\end{equation}
where $t_3$ = $t_{[OIII]}$ is in units of 10$^4$K. 
Eqs.(\ref{equation:oplus2}) - (\ref{equation:otot}) were taken from
Pagel et al. (1992), Eq.(\ref{equation:t2t3}) was taken from Garnett (1992).  
Using the value of O/H derived from Eq.(\ref{equation:ohp})
and measured line intensities Eqs.(\ref{equation:oplus2}) -
(\ref{equation:t2t3}) can be solved for $t_{[OII]}$.  The value of
$n_e$ was adopted to be equal to 100 cm$^{-3}$ for every H\,{\sc ii}
region.

The value of $t_{[OII]}$ can be also found from the following expression 
for $t_P$ = $t_3$ (\cite{pil01a})
\begin{equation}
t_{P} = \frac{R_{23} + 3.09  + 7.05 P + 2.87 P^{2}}
                       {9.90  + 11.86 P + 7.05 P^{2} - 0.583 R_{23}}  
\label{equation:tf}
\end{equation}
and Eq.(\ref{equation:t2t3}). The value of $t_2$ determined from
Eqs.(\ref{equation:t2t3}),(\ref{equation:tf}) coincides with the value
of $t_2$ derived from
Eqs.(\ref{equation:ohp}),(\ref{equation:oplus2})-(\ref{equation:t2t3})
for majority of H\,{\sc ii} regions considered. However, for H\,{\sc
ii} regions, in which the bulk of oxygen is in the O$^+$ stage, the two values
of $t_2$ are not in agreement. Eqs.(\ref{equation:ohp}),(\ref{equation:oplus2})-(\ref{equation:t2t3})
give a more realistic values of $t_{2}$ for these H\,{\sc ii} regions.

%============================================================Fig 05 zno
\begin{figure}
\resizebox{\hsize}{!}{\includegraphics[angle=0]{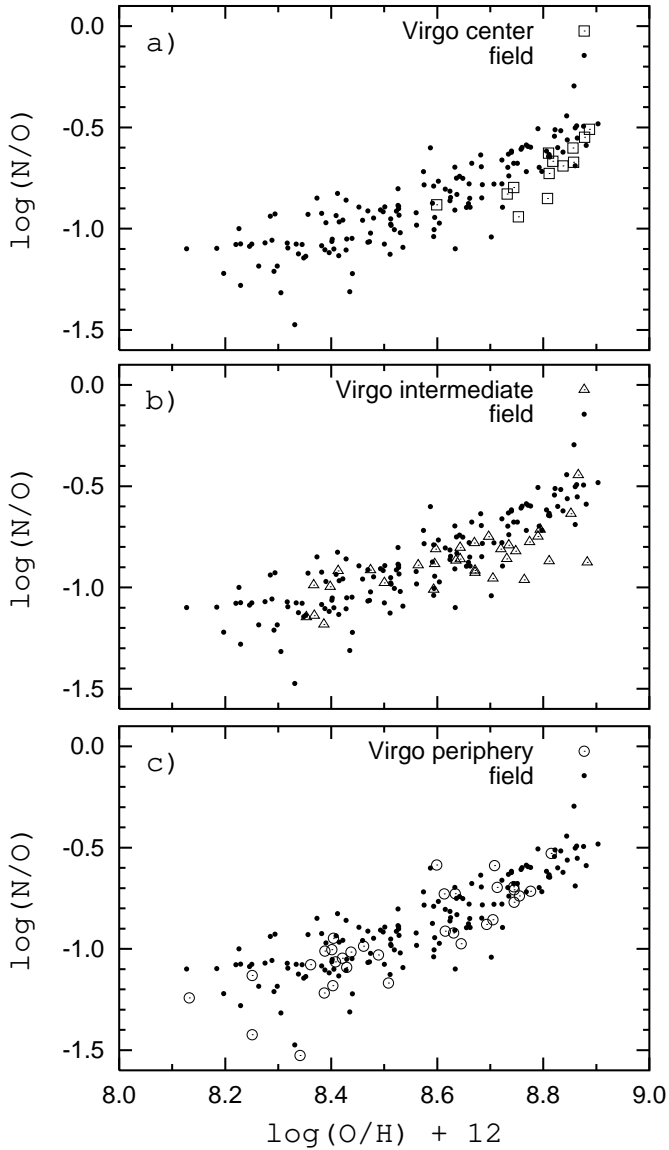}}
\caption{
N/O versus O/H diagram for Virgo and field spiral galaxies.
{\bf a.} Virgo center galaxies: NGC~4501, NGC~4571, and NGC~4689
(squares) and field spiral galaxies: NGC~628, NGC~1232, NGC~2903,
NGC~3184, NGC~3351, NGC~5194, NGC~5236, NGC~5457, and NGC~6946
(points).  {\bf b.} Virgo intermediate galaxies: NGC~4254,
NGC~4321, and NGC~4654 (triangles) and field spiral galaxies (points).
{\bf c.} Virgo periphery galaxies: NGC~4303, NGC~4651, and
NGC~4713 (circles) and field spiral galaxies (points).}
\label{figure:zno}
\end{figure}

Using Eqs.(\ref{equation:ohp}),(\ref{equation:oplus2})-(\ref{equation:t2t3})
the N/O abundance ratios were derived for all the H\,{\sc ii} regions
in Virgo and field spiral galaxies for which the [NII]$\lambda \lambda
6548, 6584$ line measurements are available in the works cited in
Table \ref{table:sample}. If only the [NII]$\lambda 6584$ line
measurement was available then the total [NII]$\lambda \lambda 6548,
6584$ line intensity was derived as I$_{[NII]\lambda \lambda 6548,
6584}$ = 1.3$\times$I$_{[NII]\lambda 6584}$.

Fig.\ref{figure:zno} shows the N/O versus O/H diagram for Virgo and field
spiral galaxies: Virgo core galaxies (NGC~4501, NGC~4571, and
NGC~4689) are presented as squares, in panel {\bf a}, the Virgo
intermediate galaxies (NGC~4254, NGC~4321, and NGC~4654) are shown as
triangles in panel {\bf b}, and the Virgo periphery galaxies
(NGC~4303, NGC~4651, and NGC~4713) are presented as circles in panel
{\bf c}. Points in all panels represent the field spiral galaxies
(NGC~628, NGC~1232, NGC~2903, NGC~3184, NGC~3351, NGC~5194, NGC~5236,
NGC~5457, and NGC~6946).

Panel {\bf a} of Fig.\ref{figure:zno} shows that the Virgo core spirals
occupy the lower part of the band outlined by the field spirals in the
N/O -- O/H diagram.  It looks as if the position of Virgo core
spirals is slightly shifted (by around 0.05 $\div$ 0.10 dex) towards
higher O/H values or towards lower N/O values relative to the overall 
position of field spirals. Is this small offset of Virgo 
core spirals in the N/O -- O/H diagram real ?  Does it reflect 
a (marginal) difference between the evolution of the Virgo core and
field spirals? It is more likely that this apparent small shift of
positions of the Virgo core spirals in the N/O -- O/H diagram is an
artifact caused by error propagation in the abundance determination 
for  H\,{\sc ii} regions in the core spirals, produced by the uncertainties 
in line measurements (only upper limit estimations of [OIII]$\lambda 5007$
line intensities are available for those H\,{\sc ii} regions), or due
to the uncertainty in the extrapolation of the P--method into an area 
of extreme values of the parameters P and R$_{23}$ (the value of P is 
equal or close to zero, Fig.\ref{figure:deva}, for all the H\,{\sc ii} 
regions of the Virgo core spirals, and the R$_{23}$ value is usually 
less than 1 while the calibration is based on the H\,{\sc ii} regions 
with R$_{23}$ $>$ 2). It should be particularly emphasized that we 
cannot discard the possibility that this
 apparent offset has an artificial origin, since it is small (around 
0.05 $\div$ 0.10 dex), comparable to the expected errors in the oxygen and 
nitrogen abundances. Therefore error statistics prevent us 
from any possible data overinterpretation; though we stress that 
more and better data are needed to disentangle the actual nature of 
the offset revealed here.

Examination of Fig.\ref{figure:zno} (panels b and c) shows that the
Virgo periphery and intermediate spiral galaxies occupy the same band
in the N/O -- O/H diagram as do field spiral galaxies, i.e. they appear
indistinguishable from field spirals in the N/O -- O/H diagram.

Thus, the comparison of N/O abundance ratios in H\,{\sc ii} regions of
the nine Virgo and nine field spiral galaxies confirms (or at least is not 
in conflict with) the above conclusion that Virgo spirals result 
indistinguishable from field spirals. If there is an actual difference 
between abundance properties of the Virgo and field spirals this difference 
should be small and appear entirely masked by the errors.

\section{Virgo Spirals and Their Multiphase Chemical Evolution Models}

Shields et al. (1991) and Skillman et al. (1996) argued that the high oxygen 
abundances in the HI deficient Virgo spirals can be explained by the fact that 
the infall of metal-poor gas onto the galaxy is inhibited in the cluster
environment, and the Virgo core spirals may have evolved more nearly
in the manner of the closed box "simple model" of chemical
evolution. Field galaxies and galaxies on the periphery of the cluster
would have their oxygen abundances depressed by infall of metal-poor
gas.  In the light of our results it can be concluded that some field
spirals have also evolved in the same way as the Virgo core spirals do. 

Another important result of the present study is the rather low values
of the central oxygen abundances in spiral galaxies: the maximum
central oxygen abundance is $\sim$ 12+log(O/H) = 9,  about 0.3 dex
higher than the solar value 12+log(O/H)$_{\odot}$ = 8.736 (Holweger 2001)
or 12+log(O/H)$_{\odot}$ = 8.69 (Prieto, Lambert \& Asplund 2001).
The multiphase multizone models shown by Ferrini et al. (1992)
predict a maximum value of oxygen abundance $12 + log (O/H) \sim 9.10$
dex.  This is indicative of that the oxygen production used in these
models is closer to reality. Therefore, one can expect that these
models can be used to test the influence of the environment, through the
effect of the infall of gas on the disks of the Virgo galaxies and
its dependence on distance to the center of the cluster.

\subsection{Multiphase Model Description}
\small
\begin{flushleft}
\begin{table}
\caption[]{ Galaxy Characteristics}
\begin{tabular}{lcccc}
\hline \noalign{\smallskip} \\ 
Galaxy & Type & Arm & R$_{\rm eff}$ & Vel$_{\rm rot,max}$ \\
 Name & & Class & (arcsec)& (km/s) \\  
\noalign{\smallskip} 
\hline 
\noalign{\smallskip} 
NGC~4303 & 4 & 9  & 62.6  & 150 \\
NGC~4321 & 4 & 12 & 111.5 & 270 \\
NGC~4501 & 3 & 9  & 77.1  & 300 \\
\hline \noalign{\smallskip}
\end{tabular}
\label{galaxies}
\end{table}
\end{flushleft}
\normalsize

The multiphase model used here was first applied to the Solar
Neighborhood (\cite{fer92}), and to the whole Galactic (MWG) disk
(\cite{fer94} -- hereafter FMPD) and bulge (\cite{mol95}). The same
model was then applied to disks (\cite{mol96}, 1999) and bulges
(\cite{mol00}) of some spiral galaxies of different morphological types
with reasonable success.  We now present the application of this
multiphase chemical evolution to three galaxies in Virgo, NGC~4501,
close to the center, an intermediate galaxy as NGC~4321 and a HI
normal galaxy NGC~4303, which is in the periphery of the cluster, as
typical examples for these different zones.

The data used in our models for these three galaxies are presented in
Table ~\ref{galaxies} where we display their morphological types (T)
and Arm Class in columns (2) and (3) as derived by Tully (1988) and
Biviano (1991), respectively.  Effective radii from  Tully (1988) are
in column (4), with the exception of NGC~4303 for which the effective
radii is from Henry et al. (1992).  The adopted distance is 16.8 Mpc.

In the multiphase model a protogalaxy is assumed to be a spheroid
composed of primordial gas with total mass M(R). For our galaxies the
masses are calculated from the rotation curves (Table ~\ref{galaxies},
column 5) obtained either via radio (\cite{guh88}) or optical
(\cite{dis90}) observations.  Each galaxy is divided into concentric
cylindrical regions 1 kpc wide. The model calculates the time
evolution of the halo and disk components belonging to each
cylindrical region. The halo gas falls into the galactic plane to form
the disk, which is a secondary structure in the multiphase framework,
formed by the gravitational accumulation of this gas at a rate which
depends on the collapse time scale $\tau_{coll}$.

In our framework an infall dependent on galactocentric distance and a
two-steps star formation law, by forming molecular clouds and then
stars, are assumed.

The infall rate $f$ is inversely proportional to the collapse time
scale $\tau_{coll}$, which is assumed to depend on galactocentric
radius, through an exponential function with a scale length $l$:
\begin{equation}
\tau_{coll}(\rm R)=\tau_{0}e^{-(\rm R-R_{0})/l}
\end{equation}

The characteristic time scale $\tau_{0}$ is defined for every galaxy
as being appropriate to a region equivalent to the Solar Neighbourhood
in MWG located at a radius $\rm R_{0}$ which is calculated from $\rm
R_{\rm eff}$.  This characteristic time scale is calculated for every
galaxy via the total mass of the galaxy.  We compute, from the
rotation curves, the total mass for each galaxy. This defines the
collapse time scale for each one of them through the expression
$\tau_{0} \propto M_{9}^{-1/2} T$ (\cite{gal84}), where $M_{9}$ is the
total mass of the galaxy in 10$^{9}M_{\odot}$ and T is its age,
assumed 13 Gyr.  Thus, we obtain $\tau_{0}$ for each galaxy, through
the proportion: $\tau_{0}=\tau_{\odot}(M_{9},gal/M_{9,MWG})^{-1/2}$,
the value of $\tau_{\odot}=4 $ Gyr having been determined in Ferrini 
et al. (1992) for the Solar Region.

%+++++++++++++++++++++++++++++++++++++     Fig 06   ngc4303
\begin{figure*}
\resizebox{1.0\hsize}{!}{\includegraphics[angle=0]{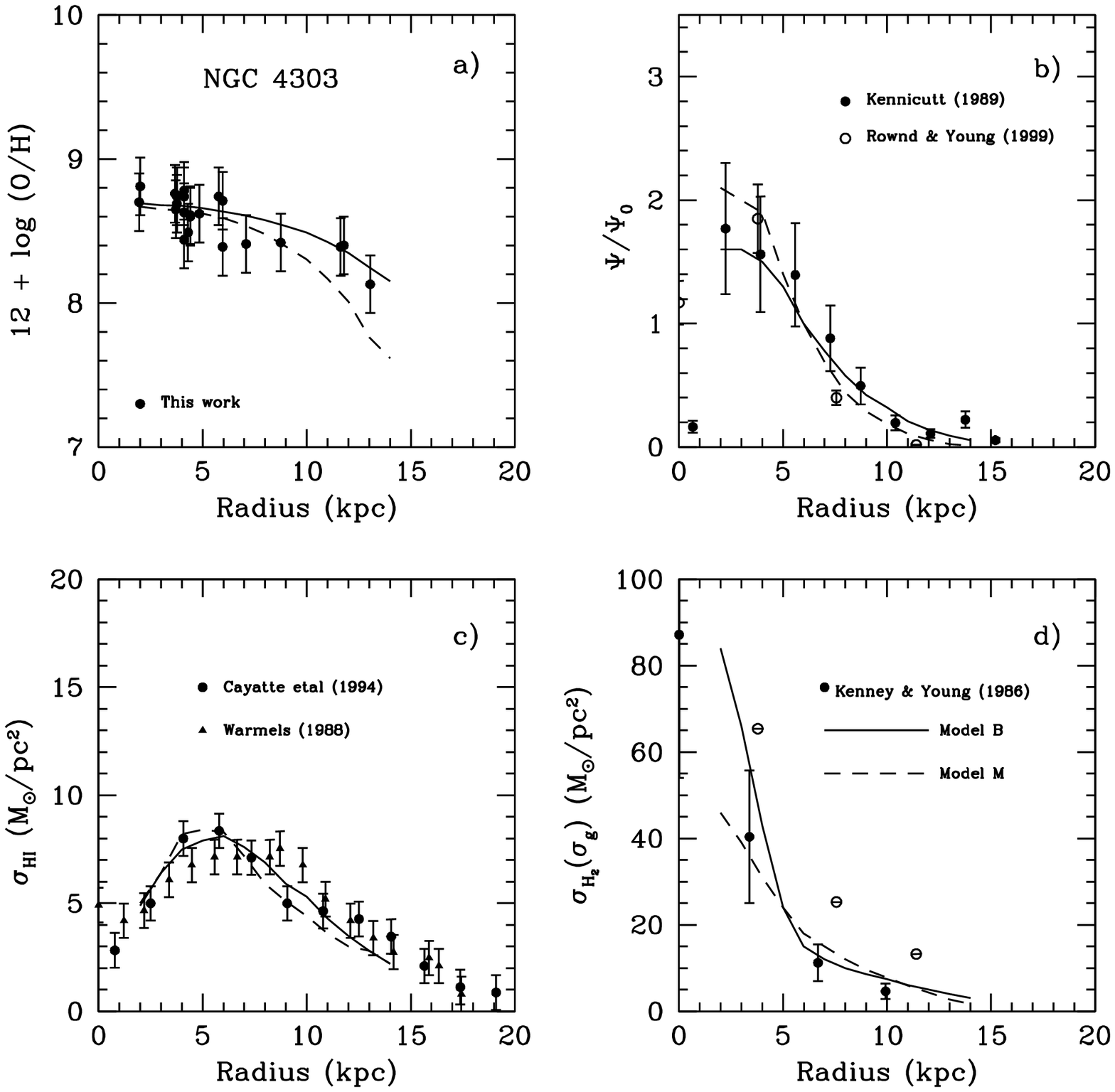}}
\caption[]{ Predicted Radial distributions at the present epoch
for NGC~4303: (a) Oxygen abundance $12+log (O/H)$; (b) Star Formation
Rate surface density normalized to the equivalent to solar region 
value$\Psi/\Psi_{0}$; (c) diffuse or atomic gas H\,{\sc i} and
(d) molecular gas H$_{2}$ surface density. Model is represented
by the lines while symbols are the observational
data}\label{ngc4303}
\end{figure*}

%+++++++++++++++++++++++++++++++++++++     Fig  07 ngc4321
\begin{figure*}
\resizebox{1.0\hsize}{!}{\includegraphics[angle=0]{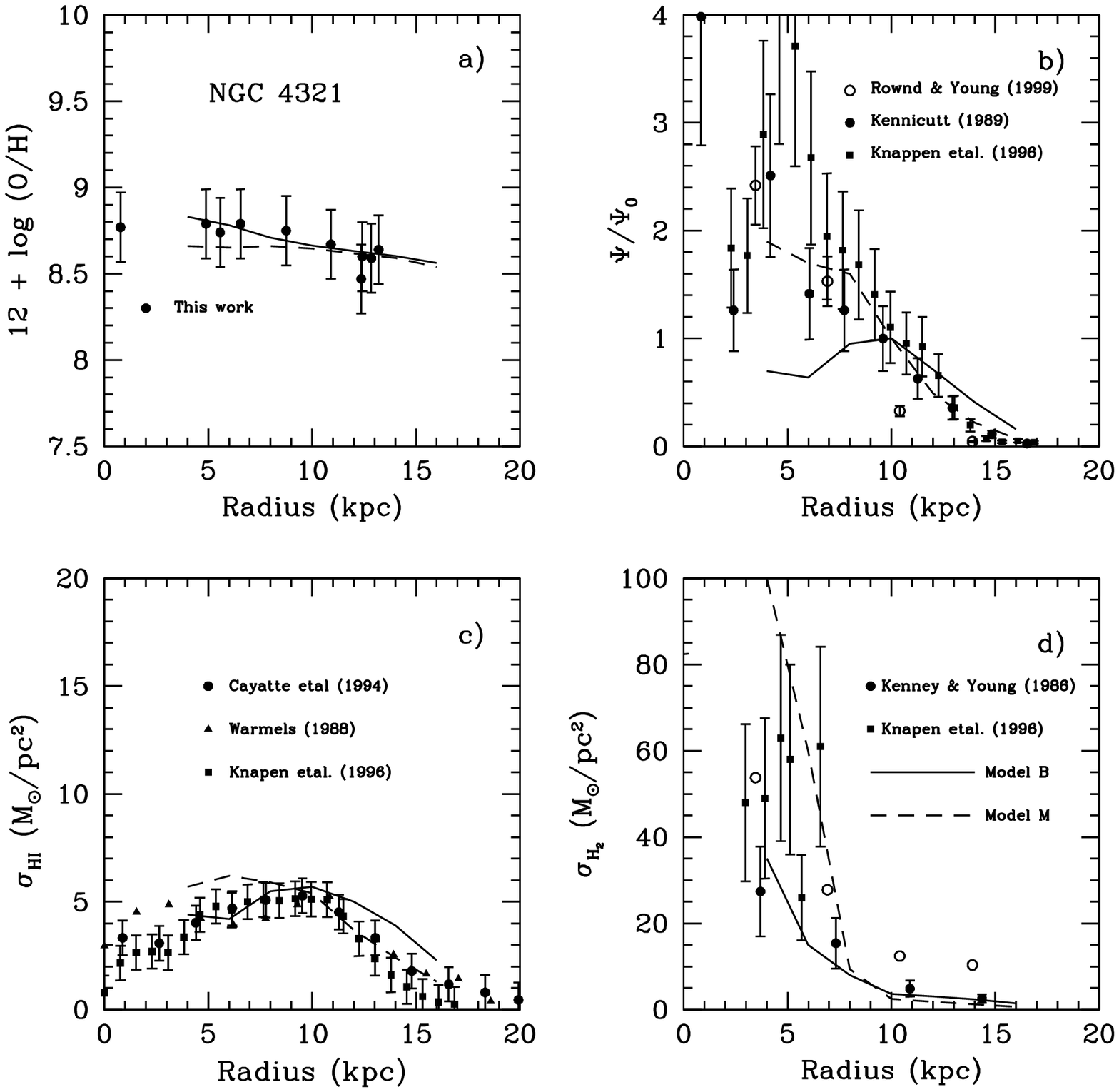}}
\caption[]{ Predicted Radial distributions at present time for
NGC~4321 with the same panels and symbols than in Fig.~\ref{ngc4303}}
\label{ngc4321}
\end{figure*}

%+++++++++++++++++++++++++++++++++++++       Fig 08 NGC4501
\begin{figure*}
\resizebox{1.0\hsize}{!}{\includegraphics[angle=0]{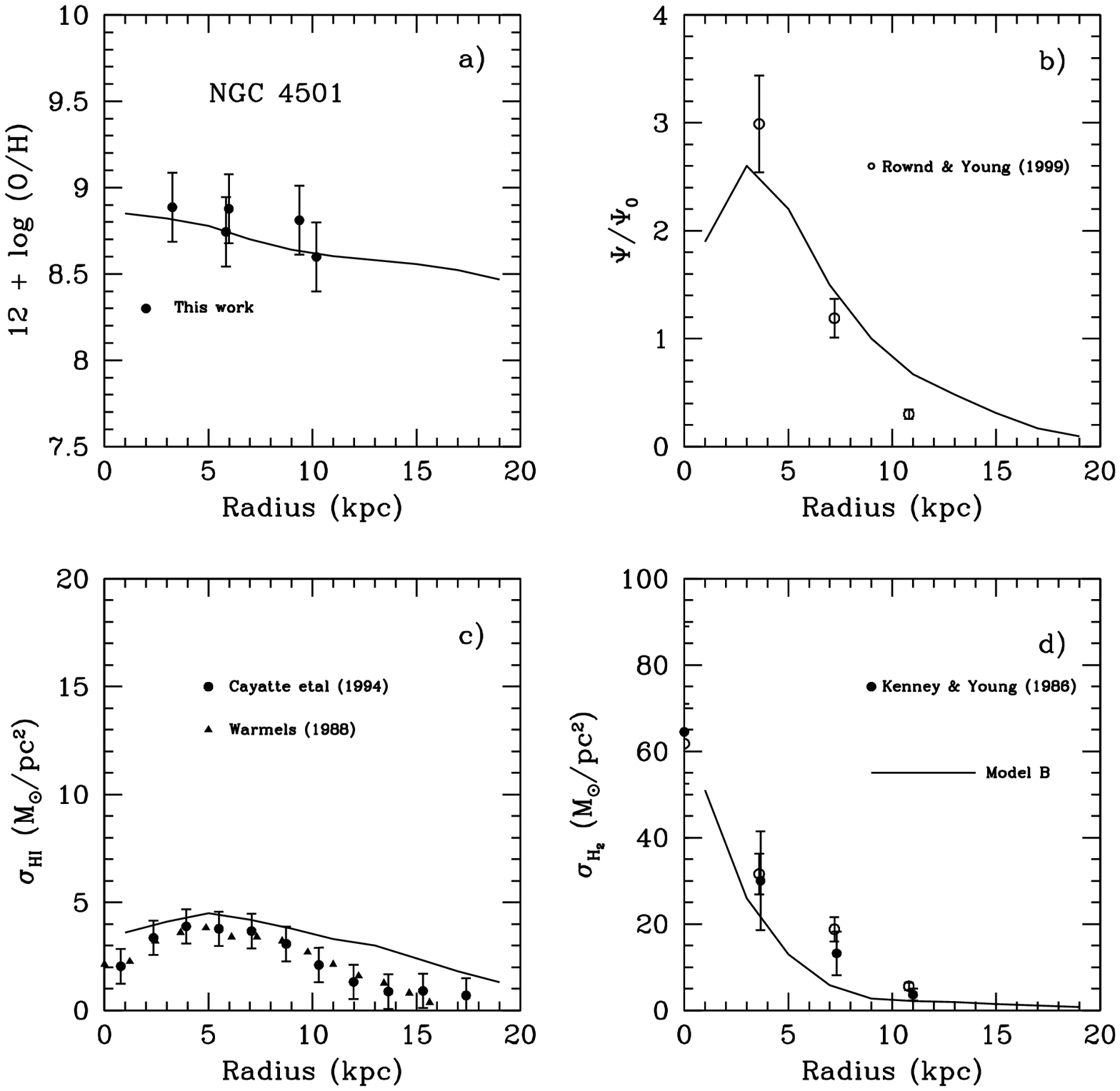}}
\caption[]{ Predicted Radial distributions at present time for
NGC~4501 with the same panels and symbols than in Fig.~\ref{ngc4303}}
\label{ngc4501}
\end{figure*}

Following the relation between collapse time scale and mass, it is
evident that $\tau_{coll}$ must be variable with the galactocentric
radius. If the total mass surface density follows an exponential law,
such as the surface brightness does for spiral disks, the resulting
$\tau_{coll}$ must increase along the radius with a scale length $ l
\propto Re$ (where $Re$ is the scale-length for the surface brightness
radial distribution). We use $l=4$ kpc but we must take into account
that $Re$ usually decreases for later type galaxies and is larger for
the earlier ones, and therefore the selection of this parameter have
some uncertainties.

In the various regions of the galaxy  we allow for different
phases of matter aggregation: diffuse gas ({\sl g}), clouds ({\sl c},
only in the disk), low-mass ($s_{1}, m < 4 M_{\odot}$) and massive
stars ($s_{2}, m \ge 4 M_{\odot}$), and remnants. The mass in the
different phases changes through several conversion process:

\begin{enumerate}
\item Star formation by the gas spontaneous fragmentation in the halo
($\propto K g_{H}^{1.5}$)
\item Formation of the disk by the accumulation of diffuse gas from
the halo ($\propto f g_{H}$)
\item Cloud formation in the disk from diffuse gas ($\propto \mu
g^{1.5}$)
\item Star formation from cloud-cloud collisions ($\propto H c^{2}$)
\item Induced star formation via massive star-cloud interactions
($\propto a c S_{2}$)
\item Diffuse gas restitution from these two star formation processes
\end{enumerate}

The rates for these processes are proportional to parameters $\mu$,
{\sl H}, {\sl a} and {\sl K}, which depend on galactocentric radius
through the equations computed in FMPD.  The proportionality factors
of those equations are the corresponding efficiencies of processes,
that is the probability of cloud formation, $\epsilon_{\mu}$, of
cloud---cloud collision, $\epsilon_{H}$, and of the interaction of
massive stars with clouds, $\epsilon_{a}$ in the disk, and the
efficiency to form stars in the halo, $\epsilon_{K}$. For each galaxy
these characteristic efficiencies must be chosen, although the term of
the induced star formation is associated to a local process and, as a
result, its coefficient $\epsilon_{a}$ is considered independent of
both position and morphological type. The last term $\epsilon_{K}$ is
also assumed constant for all halos, thus being independent of
morphological type. The other rates $\epsilon_{\mu}$ and
$\epsilon_{H}$, depend however on the Hubble type and/or the arm
class, and their adopted variation range is determined following the
arguments given by Ferrini \& Galli (1988), as discussed in 
Moll\'{a} et al. (1996, 1999).

\small
\begin{flushleft}
\begin{table}
\caption[]{Computed Models}
\begin{tabular}{lccccccccc}
\hline \noalign{\smallskip} \\ 
Galaxy & Model & R$_{0}$ &$\tau_{0}$ &$\epsilon_{\mu}$ & $\epsilon_{H}$\\
 Name  &       &  (kpc)  & (Gyr) &  &    \\  
\noalign{\smallskip} 
\hline 
\noalign{\smallskip} 
NGC~4303 & B &  6 & 8    & 0.22 & 0.008 \\ %mer3   
NGC~4303 & M &  6 & 16   & 0.20 & 0.005 \\ %mer15
NGC~4321 & B & 10 & 7.5  & 0.45 & 0.160 \\ %mer2 
NGC~4321 & M & 10 & 15   & 0.45 & 0.320 \\ %mer3 
NGC~4501 & B & 8  & 6.5  & 0.40 & 0.150 \\ %pil1
\hline \noalign{\smallskip}
\end{tabular}
\label{models}
\end{table}
\end{flushleft}
\normalsize

In order to reduce to a minimum the freedom degree due to the high
number of parameters, the model has been applied following a precise
strategy: first, it was used for the Solar Neighbourhood, followed by
the MWG model, to determine the characteristic parameters of MWG. Then
we afforded the final generalization for some other spiral galaxy, by
assuming that the two efficiencies $\epsilon_{a}$ and $\epsilon_{K}$
do not change from galaxy to galaxy.  Therefore, we allow to change
from galaxy to galaxy, only the characteristics collapse time scale,
$\tau_{0}$, depending on the total mass and the two efficiencies,
$\epsilon_{\mu}$ and $\epsilon_{H}$, following the morphological
type. The selection of each one of these input parameters in our
models will be the matter of discussion in the following subsection.

The adopted initial mass function (IMF) is taken from Ferrini et al. (1990).
The enriched material proceeds from the restitution from dying stars,
considering their nucleosynthesis, their IMF (and the delayed
restitution) and their final fate, via a quiet evolution, or Type I or
II supernova (SN) explosions.  Nucleosynthesis yields are from
Renzini \& Voli (1981), and Woosley \& Weaver (1995) for low-mass 
and intermediate, and massive stars, respectively. The type I supernova 
explosions release mostly iron
 following Nomoto et al. (1984) and Branch \& Nomoto (1986) at a slower rate,
by implying that the iron appear at least 1 Gyr later than the
so-called {\sl $\alpha$-elements} (Oxygen, Magnesium ...) ejected by
the massive stars.

\subsection{Computed models}

The collapse time scales have been chosen according to the above
paragraph and their values are shown in Table~\ref{models}.  For the 
efficiencies, they usually turn out to be larger for earlier
morphological types and lower for the later ones in MFD96, and this is
useful for a first selection of the values for a given galaxy.  The
fine tuning is then performed, once the collapse time scale
parameters $\tau_{0}$ and $l$ chosen, when $\epsilon_{\mu}$ and
$\epsilon_{H}$ are constrained to reproduce the observational data.
We use for this task the observed radial distributions for atomic gas
H\,{\sc i} (\cite{war88,cay90}), molecular gas H$_2$,
(\cite{ken88,ken89,row99}), star formation rate surface densities,
obtained from H$\alpha$ fluxes, from Kenney \& Young (1989) and 
Rownd \& Young (1999),
and oxygen abundances of the gas phase $12+ log(O/H)$ derived above. For
NGC~4321 we also used the most recent data on H${\alpha}$ fluxes and
H\,{\sc i} and H$_{2}$ radial distributions from Knapen \& Beckman (1996) and
Knapen et al. (1996).

The set of models is shown in Table~\ref{models} where we give, for
the region located at a distance $R_{0}$, column (6), equivalent to the
Solar Region, the corresponding values of the characteristic collapse
time scale, column (7) and the two efficiencies $\epsilon_{\mu}$ and
$\epsilon_{H}$, columns (8) and (9).

\subsection{The present-day radial distributions}

In the following graphs we show the results derived from the chemical
evolution models defined by the input data of Table~\ref{models},
together with the corresponding observational data.
Fig.~\ref{ngc4303}, ~\ref{ngc4321} and ~\ref{ngc4501} show for each
of the studied galaxies, the radial distribution of the oxygen
abundance in panel (a); the star formation rate surface density
normalized to its value for each equivalent solar region $\Psi_{0}$ in
(b), the gas surface density of diffuse H\,{\sc i} gas in (c), and the
gas surface density of molecular gas H$_{2}$ in (d).

%+++++++++++++++++++++++++++++++++++++         Fig 09  infall
\begin{figure}
\resizebox{1.0\hsize}{!}{\includegraphics[angle=0]{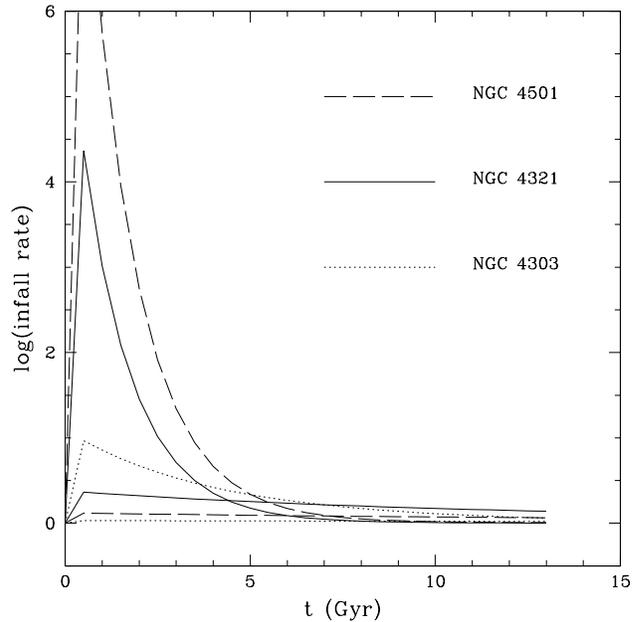}}
\caption[]{ Infall rate evolution for two regions in each galaxy, 
inner (4 kpc) -strong line-- and outer (14 kpc) --slight line--}
\label{infall}
\end{figure}

%+++++++++++++++++++++++++++++++++++++         Fig 10  muoh
\begin{figure}
\resizebox{1.0\hsize}{!}{\includegraphics[angle=0]{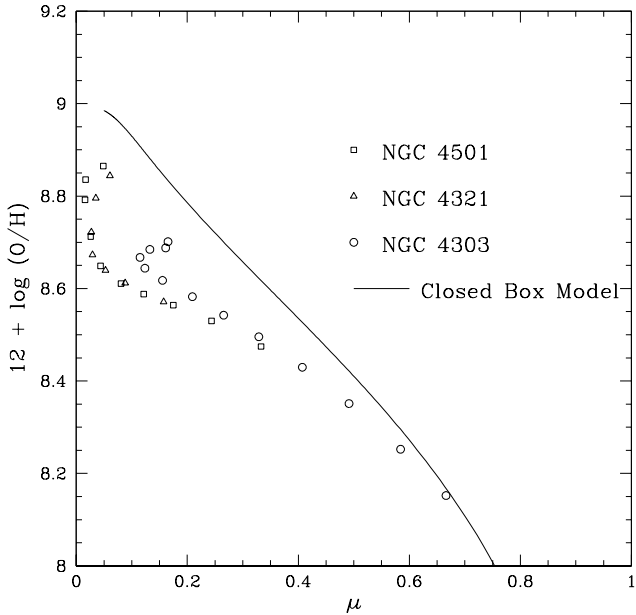}}
\caption[]{ Relation of oxygen abundances with the gas fraction $\mu$.
The solid line represent the closed box model prediction, while points
correspond to the radial distributions of oxygen abundances for the three 
galaxies at the present time as indicated in the figure}
\label{muoh}
\end{figure}

Abundances are really well reproduced for the three galaxies. By using the 
results derived in this work, the fit improves extraordinarily, 
in comparison with models from Moll\'{a} et al. (1999), where abundance 
estimates from the R$_{23}$ --method were used, thus supporting our new 
oxygen abundances.

Atomic gas radial distributions are quite well reproduced for all
three galaxies, although they have a high gas surface density in the
outer disks in comparison with the data. This difference between data and
models does not appear in NGC~4303 models.  For Model~B of NGC~4321 it is
$\sim 1-1.5 M_{\odot}/pc^{2}$ for $R > 10 kpc$, while for Model~M
it does not exist. This implies that this last model is better able 
reproduce the outer disk, while the first one better fits the data in
the inner regions.  For the NGC~4501 model, the difference between data
and models appears at $R> 8 kpc$ and takes values comparable to 
the mean density in this outer disk. Summarizing, the discrepancy
between models and data increases with decreasing distances to the
center of the cluster, simulating perfectly the behavoir of the
deficiency of the gas.  This observed depletion of the diffuse gas
from the outer regions of disks needs some mechanism producing this
gas loss, which is not included in our models. Therefore we cannot
reproduce this special feature of H\,{\sc i} deficiency in the Virgo
outer disks.

Molecular gas radial distributions differ slightly in the inner
regions, but always close to the error bars.  The uncertainties in the
estimation of the molecular cloud masses, which are obtained from CO
intensities through a calibration parameter $\chi$, are large because
the dependence of $\chi$ on metallicity or galactocentric distance
(\cite{ver95,wil95}) in each galaxy is not completely well
determinated.

These good results in the predicted gas radial distributions
are, in fact, the consequence of searching models fitting these two
distributions. Therefore the largest differences for a galaxy may
only appear in the oxygen abundance and star formation rate radial
distributions. 

As we have already pointed out, abundances show a good behaviour, consistent 
with the data, while the shape of the star formation rate surface density 
radial distributions are well fit, showing exponential shapes, with the
exception of Model B for NGC~4321 (panel d, Fig.~\ref{ngc4321}).  This
model B predicts a star formation rate lower than the data in the inner
regions.  This also appears in Fig.~\ref{ngc4501} where a maximum rate
is produced at 3-4 kpc and it implies that the model has a strong
evolution, maybe more rapid than the actual evolution.

Thus, it seems that, following Figs.~\ref{ngc4303},\ref{ngc4321}, \ref{ngc4501}, 
our models reproduce the abundance data in Virgo galaxies well, and therefore
we can use these models to explain the time evolution of Virgo galaxies.
In order to see clearly the effects of the time evolution we show in
Fig.\ref{infall} the infall rate for two regions, located at 4 and 14
kpc, and for model B of the three galaxies. We see that
the infall rate is clearly higher in the inner region of NGC~4501 
compared to the two other galaxies, NGC~4321 rests in middle of
the two others, while NGC~4303 has a slow infall even in this central
region.  For the outer regions, the infall rate continues to be larger
for NGC~4501, intermediate for NGC~4321 and low for NGC~4303,
following the same trend as in the 4 kpc radial zone, but differences
among galaxies are now smaller. All zones and galaxies show a similar
and low infall rate at the present time. The most important 
difference resides in the fact that the infall rate had a very strong
maximum during the first two Gyr for the inner diks in both NGC~4501 and
NGC~4321, which finished at $\sim 5$ Gyr, while NGC~4303 maintains a slow
evolution, with an infall rate higher than that of the two others
from 5 Gyr up to now.

These different infall rate histories have important effects on the
oxygen abundances obtained along the galactic disk for these galaxies.
We show in Fig.\ref{muoh} the oxygen abundances, as 12 + log (O/H),
for each radial region and galaxy {\sl vs} the gas fraction
$\mu=M_{gas}/M_{disk}$. In this graph the solid line represents the
closed box model prediction, while points correspond to the three disk
radial distributions.  The positions of NGC4303 (circles), NGC4321
(triangles), and NGC4501 (squares) in the $\mu$ - O/H diagram show
that these galaxies evolve with some gas infall. We may see that the
outer regions of NGC~4303 are close to the closed box model line,
simulating a galaxy without mass exchange with the environment, while
the inner regions have abundances smaller than the ones expected from their 
gas fraction according to the closed box model.  NGC~4321 and NGC~4501 show
a strong decrease in its abundances for their inner regions, and
there is, even for the outer ones, some dilution effect.  All these
regions have very low gas fractions, which according to the closed box model
should correspond to the highest abundances, 9.0-9.10; however this is not 
the case, and the abundances are lower than these values.

Thus, our models show that both H\,{\sc i} normal and H\,{\sc i} deficient 
galaxies have their oxygen abundances depressed by infall of metal-poor gas. 
Indded, the model for the NGC~4501 (a representative of the H\,{\sc i} 
deficient Virgo core spirals) shows that the inner regions of NGC~4501 
have very low gas fractions and therefore even a slow infall rate at the 
present epoch results in a dilution effect, and the evolution of Virgo core 
spirals are somehow similar to that of field spirals. Taking into account 
that our models reproduce the data well for the three galaxies, in 
particular the gas densities and abundances, this
conclusion does not depend on the goodness of the hypotheses or
theoretical assumptions.

\section{Discussion}

In order to make sure that our models are in agreement with the whole 
set of new results for the cluster galaxies and in particular for Virgo, we 
will summarize these observations. Solanes et al. (2001) have derived the 
atomic content for a total of 1900 galaxies in 18 nearby clusters, 
aiming to study possible connections between the gas deficiency and
the characteristics of the galaxies or their environment. With these
data it is well established that the proportion of gas-poor galaxies
increases continuously toward the center of the cluster for a large
number of them. Furthermore, there are clear indications that the
removal of H\,{\sc i} occurs mostly in the outer regions of disks,
thus implying that the H\,{\sc i} disk sizes are reduced
(\cite{war88,cay90,cay94}), and that the degree of the H\,{\sc i}
depletion is related to the morphological type, early types and dwarfs
being more easily emptied of gas than intermediate types.

On the other hand, the analysis of the spectroscopic catalog of distant
clusters performed by Dressler et al. (1999) and Poggianti et al. (1999)
show that the star formation
rate has been quenched, rather than enhanced, in these galaxies. This
conclusion has been also reached by Caldwell et al. (1999), who found that 
star formation takes place in the inner regions of these
galaxies, but it is suppressed in the outer ones. Moreover the colors
of these galaxies (\cite{gav91,gav98}) do not differ significantly from
those of field galaxies for the same morphological types, and the
molecular gas content, usually more concentrated in the inner disks,
show normal masses (\cite{ken88}, 1989).

Thus, the scenario necessary to reproduce all observations must be
able to eliminate the gas of the outer regions, but not in the inner
disk, and at the same time, it has to form stars at the early times
of evolution, in order to obtain the red colors observed and the
star formation rates estimated for the present time. The most extended
idea is that the mechanism producing this removal of gas is the ram
pressure stripping due to the orbit of each galaxy around the core of
the cluster.  Under this assumption the low density diffuse gas may be
eliminated when the galaxy passes close to this core, while the higher
density molecular clouds located in the inner disk should not be
affected.

Recently, Vollmer et al. (2001) have performed some N-body simulations in
order to study the effects of this mechanism on a galaxy suffering ram
pressure stripping by its motion within the cluster. They find
that the movement of the galaxy around the cluster center provokes an
acceleration of the clouds located on the surface of the disk, which,
in turn, produces a high surface density in the center of the
galaxy. Thus, the star formation can suffer an enhancement in the inner
disk. After this phase, the gas of the outer regions is removed and,
without this fresh gas reservoir, the star formation decreases.

Within this scenario, the high density produced in the inner regions
of the galaxy is the consequence of a strong infall rate on a short
time scale. This is exactly what happens in the NGC~4501 model, where
the inner disk suffers a large infall rate in a short time, by
creating stars very rapidly. The consequence is the consumption of the
gas, and therefore star formation stops. For a Virgo periphery galaxy, 
as NGC~4303, the infall is much slower, maintaining a more constant star
formation rate along its evolution, similar to the typical one of a field
galaxy. The model for NGC~4321 has a behavior intermediate between
both of them. This explains why the star formation rate is, for the
present time, higher for NGC~4303, which still has enough gas, while
this rate has decreased for NGC~4321 from its maximum value, and it is
very depressed for NGC~4501, which has almost no gas. These model
predictions are supported by the observations from Koopmann \& Kenney 
(1998) and Rownd \& Young (1999) who find that the star formation 
efficiency decreases with increasing H\,{\sc i} deficiency among 
the Virgo cluster galaxies. In fact the H$\alpha$ luminosities give 
values $log L=$ 8.38, 7.97 and 7.96 $L_{\odot}$ for NGC~4303, 4321 
and 4501, respectively, following these latter authors.

We would like to add some other considerations about our models. Koopmann 
\& Kenney (1998) found that the morphological classification for the
Virgo cluster galaxies does not follow the same scheme as the one
used for the field galaxies. In this last case the concentration index
(measured as the bulge to disk luminosity ratio) is correlated with
the star formation in the arms, while in the cluster galaxies it
does not occur: due to the H\,{\sc i} deficiency in the outer disk,
the concentration index increases for a given H$\alpha$ flux, as a
measure of the star formation rate. Thus if it corresponds to a Sc galaxy,
the concentration index would define the galaxy as Sa. This fact
implies that the Virgo galaxies would be classified as earlier types 
whereas their star formation rates would correspond to later types of galaxies. 
In this sense, it appears remarkable that our models reproduced the observed 
characteristics of the three selected galaxies with efficiency values 
lower than those typically used for their morphological types, and closer 
to the values corresponding to one unity later morphological types.

\section{Conclusions}

The oxygen and nitrogen abundances in H\,{\sc ii} regions in the nine
Virgo spirals of the sample from \cite{ski96} and in nine selected field spiral
galaxies were re-determined with the recently suggested P -- method. 

It has been confirmed that there is an abundance segregation in the 
sample of Virgo spirals in the sense that the H\,{\sc i} deficient
Virgo spirals near the core of the cluster have higher oxygen
abundances in comparison to the spirals at the periphery of the Virgo
cluster. In general, both Virgo periphery and Virgo core spirals have
counterparts among field spirals.  
Some field spirals have a small H\,{\sc i} to optical radius ratios, 
similar to the one in H\,{\sc i} deficient Virgo core spirals. 
It has been concluded that if there is an actual difference in abundance 
properties between the Virgo and field spirals this
difference should be small or entirely masked by the errors.

These abundance results have been analyzed with the multiphase chemical 
evolution model, applied to three galaxies NGC~4501, 4321 and 4303,  considered as
typical examples of galaxies located at the center, at intermediate
locations, and at the periphery of the cluster. These models are shown
to be able to reproduce  the observations of abundances, star formation 
rate and diffuse and molecular gas densities. 
Following these models the infall rate was strong at early times and
it is very low now for the Virgo core galaxies, but with features continuously
smoother for the periphery galaxies. The consequence of the infall of
the gas is the dilution of the elements, and therefore lower abundances
than those expected with the closed box model, for the gas fractions in the
center galaxies, are predicted. Thus, the final abundances turn out to
be similar to the ones present in field galaxies.

The scenario of the ram pressure stripping usually claimed to produce
the observed characteristics in the cluster galaxies, such as a H\,{\sc i} 
deficiency, which is larger for the galaxies closer to the cluster, was  
recently simulated with N-body models by \cite{vol01}. 
Following this work, a large infall rate is produced in the inner 
regions of the disk of these galaxies, as a consequence of their movement 
towards the cluster core, with a posterior phase of quenching of the star
formation; these results appear consistent with our conclusions.

\begin{acknowledgements}
We thank Prof. A.I.~D\'{\i}az and Dr. M.~Castellanos for useful discussions. 
L.S.P. is grateful to the staff of the Grupo de Astrof\'{\i}sica 
(Universidad Aut\'onoma de Madrid) for hospitality during a visit when a 
part of this work was carried out. 
This study was partly supported by the NATO grant PST.CLG.976036 (L.S.P., M.M., 
and F.F.), the Joint Research Project between Eastern Europe and Switzerland 
(SCOPE) No. 7UKPJ62178 (L.S.P.), the grant No 0312 of the Ukrainian fund for 
fundamental investigations (L.S.P.), and the Spanish Ministerio de Ciencia y 
Tecnologia project AYA -- 2000 -- 093 (M.M.).
We thank an anonymous referee for helpful comments.
\end{acknowledgements}

\end{document}